\pdfoutput=1
\documentclass{article}
\usepackage[T1]{fontenc}
\usepackage[ruled,linesnumbered]{algorithm2e}
\usepackage{amsmath}
\usepackage{amssymb}
\usepackage{graphicx}
\usepackage{hyperref}

\SetAlFnt{\small}
\SetAlCapFnt{\small}
\SetAlCapNameFnt{\small}
\SetAlCapHSkip{0pt}
\IncMargin{-\parindent}

\begin{document}

\markboth{M. P. Startek}{An asymptotically optimal, online algorithm for weighted random sampling with replacement}

\title{An asymptotically optimal, online algorithm for weighted random sampling with replacement}
\author{MICHA{\L}\ PIOTR STARTEK\\
{\small University of Warsaw, Faculty of Mathematics, Informatics, and Mechanics}
}
\maketitle

\begin{abstract}
This paper presents a novel algorithm solving the classic problem of generating a random sample of size $s$
from population of size $n$ with non-uniform probabilities. The sampling is done with replacement. 
The algorithm requires constant additional memory, and works in $\mathcal{O}(n)$ time (even when $s >> n$, in which case 
the algorithm produces a list containing, for every population member, the number of times it has been selected
for sample). The algorithm works online, and as such is well-suited to processing streams. 
In addition, a novel method 
of mass-sampling from any discrete distribution using the algorithm is presented.
\end{abstract}

{\let\thefootnote\relax\footnote{
Author's address: M. Startek \\
Wydzia{\l}\ Matematyki, Informatyki i Mechaniki\\
Uniwersytetu Warszawskiego\\
ul. Banacha 2\\
02-097 Warsaw\\
Poland
}}

\renewcommand{\O}{\mathcal{O}}
\newcommand{\iid}{\mathrm{iid}}
\newcommand{\U}{\mathcal{U}}
\newcommand{\UI}{\U(0, 1)}
\newcommand{\iidfrom}{\underset{\iid}{\sim}}
\newcommand{\Beta}{Beta}

\section{Introduction}
Assume that we are given a population of elements $P = \{e_i\}_{i=0}^{n-1}$, $n \in \mathbb{N}$ (at least at first, the problem of infinite populations will be elaborated on later),
along with a sequence of probabilities of each element of $P$, denoted $\{p_i\}_{i=0}^{n-1}$, such that $\sum_{i=0}^{n-1} p_i = 1$, and a single number $s$, the sample size, 
which might be greater or lower than $n$. The task is to compute a random sample of size $s$ from the population $P$, such that each element $X_i$ from the sample
is one of the elements of $P$, each with its corresponding probability. Note that without loss of generality we can (and will) assume that $P = \{0, ..., n-1\}$

The algorithm assumes a non-naive (constant-time) implementation of procedures for sampling single random numbers from the 
beta (in the easy case, where $\alpha$ and $\beta$ parameters are integer and $\geq 1$), and binomial distributions, as well as lack of numerical errors. 
Some consideration to mitigating the effects of numerical inaccuracies will be given in later sections.

The algorithm is best presented (as the author feels) by starting from the naive algorithm, and iteratively refining it, until the desired time and memory complexity are reached.

\section{The naive algorithm}
The naive algorithm (which, despite its non-optimal costs, in practice is reasonably efficient, and is used, in its second variant, 
for example by the \emph{numpy} numeric library for Python) is based on a cruicial idea,
which will be used also in the novel version presented here. The idea is based on a geomertical intuition: if an interval $[0, 1]$ is divided into parts with lengths $p_i$,
then sampling a random number $X$ from the uniform distribution $\mathcal{U}(0, 1)$ and picking the subinterval of $[0, 1]$ into which it falls (and the corresponding element of $P$)
results in a choice of a single element of $P$ with the desired probability distribution. Efficient finding of the selected subinterval is faciliated by precomputing
an array of cumulative sums of probabilities, then performing a binary search on it. 

\begin{algorithm}[t]
\KwIn{The sequence of probabilities ${p_i}_{i=0}^{n-1}$, desired sample size $s$}
\KwOut{A multiset $R$ containing the random sample}
Compute array $\{S_i\}_{i=0}^{n-1} = \sum_{j=0}^{i-1} p_j$\;
$R = $ new empty multiset\;
\Repeat(\ $s$ times){}{
	Randomize $X \sim \mathcal{U}(0, 1)$ \;
	Find greatest $k$ s.t. $S[k] < X$ using binary search \;
	Add $k$ to $R$
      }
$R$ contains the result \;
\caption{The naive sampling algorithm}
\label{alg:one}
\end{algorithm}

The algorithm consumes $\O(n)$ time for initialization (lines 1 and 2), then $\O(s \log(n))$ time for the actual sampling, and $\O(n)$ memory space for additional data structures (not 
counting the $\O(s)$ for the result).

\section{Omitting the computation of cumulative sums table}
The first modification of the algorithm makes it possible to skip the necessity to precompute the array of cumulative sums in line one. Instead it samples
all the necessary random numbers $X_i \iidfrom \UI$, sorts the $X$ array, and then processes the $p_i$ sequence at the same time as $X$, in fasion similar to
the merge step of the mergesort algorithm.

\begin{algorithm}[t]
\KwIn{The sequence of probabilities ${p_i}_{i=0}^{n-1}$, desired sample size $s$}
\KwOut{A multiset $R$ containing the random sample}
$R = $ new empty multiset\;
$idx = 0$ \;
$cumulativeProbSum  = 0.0$ \;
Randomize $X \sim \UI^s$\;
Sort $X$ in ascending order\;
\ForEach{$x \in X$}
{
	\While{cumulativeProbSum $<$ x}
	{
		$cumulativeProbSum += p[idx]$ \;
		$idx++$
	}
	Add $idx-1$ to $R$\;
}
$R$ contains the result \;
\caption{The sampling algorithm without cumulative sums table}
\label{alg:two}
\end{algorithm}

The algorithm runs in $\O(s \log(s) + n)$ time, (which is not an improvement over the previous version): lines 4 and 5 take a
total of $\O(s \log(s))$, the outer loop runs $s$ times, while the inner loop runs a total of $n$ times (as the variable $idx$ is bounded by $n$).
The algorithm uses $\O(s)$ memory.

\section{Omitting the sorting}

The algorithm might be further improved if the table $X$ could be generated in an already sorted order. This is, in fact, possible: it is a well-known fact
that if $X_0, ..., X_{n-1} \iidfrom \UI$, then $min(X_0, ..., X_{n-1}) \sim \Beta(1,n)$, and is the first element of 
the sought table~\cite{beta}. Since the variables are independent, then, after sampling the minumum
using this method, it is easy to see that the remaining variables (under condition that they have to be not less than the minimum) are distributed according to
$\U(M, 1)$, where $M$ is the minimum. The second-lowest variable might be sampled with the same method after rescaling $\U(M, 1)$ to $\U(0, 1)$, and so on.

In fact, this allows us to drop the step of precomputing the table $X$ altogether, and to just compute the consecutive variables "on the go", making the algorithm 
capable of online operation, as well as improving the runtime.

\begin{algorithm}[t]
\KwIn{The sequence of probabilities ${p_i}_{i=0}^{n-1}$, desired sample size $s$}
\KwOut{A multiset $R$ containing the random sample}
$R = $ new empty multiset\;
$idx = 0$ \;
$cumulativeProbSum  = 0.0$ \;
$currentX = 0.0$ \;
\For{$i$ in $s, ..., 1$}
{
	$currentX += \Beta(1, i) * (1.0 - currentX)$ \;
	\While{$cumulativeProbSum < currentX$}
	{
		$cumulativeProbSum += p[idx]$ \;
		$idx++$ \;
	}
	Add $idx-1$ to $R$ \;
}
$R$ contains the result \;
\caption{The online algorithm}
\label{alg:three}
\end{algorithm}

The algorithm runs in $\O(n+s)$ time, and requires constant additional memory if working online: in that case, every intermediate result is immediately provided
to the calling procedure for consumption (and possibly, immediately discarded), instead of being explicitly stored in $R$.

\section{The case of $s >> n$}

The practical speed of the above algorithm is constrained by the speed of the sampling from the beta distribution, the remaining operations being trivial in comparison.
This provides an opportunity for optimization: if the population in small with respect to the numer of samples required, then the algorithm will have to sample from the beta
distribution many times for any population member. This can be avoided by changing the reasoning: insead of asking "where will the next $X_i$ be?" we can ask "how many $X\mathrm{es}$
we will encounter while going through the current $p_i$?". The answer to that, for $p_1$ is the binomial distribution: $Binom(s, p_1)$. For further $p_i\mathrm{s}$ 
the answer is the same distribution, only conditioned on the number of $X\mathrm{es}$ and the population probability already consumed: $Binom(s-|X_{used}|, p_i / (1.0 - \sum_{j=0}^{i-1}p_j))$.This is in fact a standard algorithm for sampling from multinomial distribution (which is exactly the same problem as random sample with replacement: only the former terminology
is most often used in contexts where $s >> n$, and the latter otherwise).

\begin{algorithm}[t]
\KwIn{The sequence of probabilities ${p_i}_{i=0}^{n-1}$, desired sample size $s$}
\KwOut{A multiset $R$ containing the random sample}
$R = $ new empty multiset\;
$idx = 0$ \;
$cumulativeProbSum  = 0.0$ \;
\For{$i$ in $0, ..., n-1$}
{
	Randomize $N \sim Binom(s, p[i] / (1.0 - cumulativeProbSum))$ \;
	Add $i$ to $R$ with multiplicity $N$ \;
	$s -= N$ \;
	$cumulativeProbSum += p[i]$ \;
}
$R$ contains the result \;
\caption{The online algorithm for $s > n$}
\label{alg:four}
\end{algorithm}

The provided algorithm runs in $\O(n)$ time (assuming that $R$ behaves like a counter, and increasing the count of a given item is done in constant time), and in constant memory.
It is capable of working online. It should be noted, that although it achieves the optimal theoretical asymptotic runtime, its practical implementation will be very inefficient when $s << n$:
even if a sample consisting only of one element is desired, it will perform $n$ expensive operations of sampling from binomial distribution.

\section{A practical algorithm}
The previous two algorithms are opposites in terms of their practical pessimistic case: the one using beta distribution has to randomize once per each requested sample member, and so, runs
fast if $s << n$, and slowly if $n << s$, while the one using the binomial distibution has to randomize once for every member of the population, 
and as such is efficient in practice only for large values of $s$ and small $n$. It turns out that it is actually possible to create a hybrid algorithm which combines the strengths of 
both of them. What's more, the algorithm doesn't work by first examining the data, and then choosing one of the previous versions and runnig it, instead it adapts "on the fly", is capable of
switching back and forth between modes during runtime as needed, and does not need to examine the data in advance, which keeps it compatible with online operation.

Recall the metaphor of a segment divided into fragments corresponding to the population members, with lengths equal to their probabilities. The algorithm may be imagined as
if walking along the segment, picking its sample along the way. It can make two kinds of steps: first is the "beta" step, with constant average length, which may pass over multiple 
small population elements, and results in adding to the sample the population member in which it ends (with multiplicity of one). The disadvantage is that if a 
large population member is encountered then it may take multiple beta steps to pass it. The other kind of step, the binomial step immediately travels forward to the end
of the current population member, adding it to the sample with multiplicity according to the result of randomization. Obviously it makes sense to use this type of step 
while traversing population members with large probabilities. This is achieved through the condition in line 7: the expected number of samples randomized from the current 
member of the population is compared to a constant (1.0 here). The result of this comparison is used to determine whether to proceed in beta or in binomial mode.
Any positive constant here is good enough to achieve the theoretical bounds, however, in practice it
is best to choose it based on the relative costs of sampling from beta and from binomial distributions.

The algorithm is still online (although presented in non-online form for readability), works in constant memory if online (results are immediately consumed by caller, 
instead of being stored in $R$). A careful reader might notice that the algoritm as presented runs in pessimistic $\O(n+s)$ time. The pessimistic time is achieved if 
the algorithm encounters an element of
the population with probability small enough that it decides to use the beta mode, 
but, due to bad luck, proceeds to draw $\O(s)$ infinitesimal samples from the beta
distribution before leaving the element and proceeding forward. This may be easily 
avoided by adding a hard condition that would force a binomial mode after a constant number
of consecutive beta samples. This was omitted from the main code presented here for 
readability, and also because it is not a concern for any practical application. 
However, with the hard limit, the algorithm
can perform at most $n$ binomial samples (as each binomial sample increases the $idx$ 
variable - which is bounded by $n$ - and with the hard limit it is possible to perform
at most $\O(n)$ beta samples) - therefore its runtime is bounded by $\O(n)$.

\begin{algorithm}[t]
\KwIn{The sequence of probabilities ${p_i}_{i=0}^{n-1}$, desired sample size $s$}
\KwOut{A multiset $R$ containing the random sample}
$R = $ new empty multiset\;
$idx = 0$ \;
$cumulativeProbSum  = 0.0$ \;
$currentPosition = 0.0$ \;
\While{$s > 0$}
{
	$cumulativeProbSum += p[idx]$ \;
	\While{$(cumulativeProbSum - currentPosition) * s / (1.0 - currentPosition) < 1.0$}
	{
		$currentPosition += \Beta(1.0, s) * (1.0 - currentPosition)$ \;
		\While{$cumulativeProbSum < currentPosition$}
		{
			$idx++$ \;
			$cumulativeProbSum += p[idx]$ \;
		}
		Increase multiplicity of $idx$ in $R$ by $1$ \;
		$s -= 1$ \;
		\If{$s == 0$}{Terminate algorithm, $R$ contains the result}
	}
	Randomize $N \sim Binom(s, (cumulativeProbSum - currentPosition)/(1.0-currentPosition))$ \;
	Increase multiplicity of $idx$ in $R$ by $N$ \;
	$s -= N$ \;
	$idx += 1$ \;
	$currentPosition = cumulativeProbSum$ \;
}
$R$ contains the result \;
\caption{The final algorithm}
\label{alg:five}
\end{algorithm}

\section{Practical notes}

The algorithms presented here depend heavily on good implementations of functions for sampling from binomial and from beta distribution.

In particular, anyone undertaking the implementation of the algorithm is advised to write a custom version of function for sampling from $\beta(\alpha, \beta)$ distribution:
the algorithm always samples with $\alpha = 1$, and in this case, the distribution has an explicit, invertible CDF - and so, a custom sampler using the inverse CDF 
method will always be faster than a sampler from any scientific library which has to handle the general case.

Regarding the binomial distribution, the C++11 function for sampling as implemented by the GNU project's libstdc++ (standard C++ library on most Unix systems) is inadequate
for the task as it seems to have non-constant complexity with regard to its parameters. In any tests performed I have used the function as implemented by the Boost
library which seems to work not only faster than libstdc++'s, but also runs in constant time. 

Regarding the numerical stability of the algorithm, the algorithm computes a cumulative sum of all encountered probabilities, which is of course tricky. If precise numerical
correctness is required, then the summation should be done using Kahan's~\cite{kahan} or even Shewchuk's~\cite{shewchuk} summation algorithms. However, for most if not all 
practical purposes imaginable, this is not necessary. Special care must be taken, however, as sometimes, due to numerical errors, the last sampling 
from binomial distribution might be performed with probability of success slightly greater than 1 - the programmer must ensure that the sampling function just assumes 1 instead of
crashing.

As a side note: the algorithm does benefit slightly from having its input data sorted (in either ascending or descending order, doesn't matter): it casues the algorithm to perform less
switches between beta and binomial mode, and so, it minimises the number of elements which are partially dealt with in binomial mode, and partially in beta mode. The very slight
speed benefit does not justify spending the computational time (and especially the loss of asymptotic optimality and the ability to work online) needed to sort the data.
However, this means that in order to avoid any bias in runtime tests described further, the input data for all tests was randomly shuffled.

\section{Comparison with other sampling methods and runtime tests}
\begin{table}
\begin{tabular}{|l|r|r|r|r|}
\hline
 Algorithm             & Pessimistic runtime & Additional memory & Calls to RNG & Can work online \\
                       &                     & used              &              &   \\
\hline
 Algorithm~\ref{alg:one} & $\O(n + s \log n)$ & $\O(n)$          & $\O(s)$       & No              \\
 (numpy)                 & &  & & \\
\hline
Algorithm~\ref{alg:two} & $\O(n + s \log s)$ &  $\O(s)$          & $\O(s)$ & Yes             \\
\hline
Algorithm~\ref{alg:three} & $\O(n + s)$ &  $\O(1)$          & $\O(s)$ & Yes           \\
\hline
Algorithm~\ref{alg:four} & $\O(n)$ &  $\O(1)$          & $\O(n)$ & Yes           \\
\hline
Algorithm~\ref{alg:five} & $\O(n)$ &  $\O(1)$          & $\O(\min(n, s)) $ & Yes           \\
\hline
Walker's algorithm & $\O(n+s)$ &  $\O(n)$          & $\O(s) $ & No           \\
\hline
\end{tabular}

\label{tbl-the}
\caption{A comparison of properties of sampling algorithms}
\end{table}
\begin{figure}
\begin{center} Comparison of runtimes \end{center}
\begin{tabular}{|c|c|}
\hline
\multicolumn{2}{|c|}{\includegraphics[width=0.8\linewidth,trim={0 640 0 0},clip]{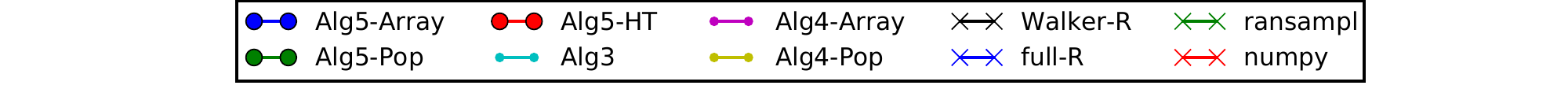}} \\
\hline
\multicolumn{2}{|c|}{"Gaussian" population} \\
\hline
\includegraphics[width=0.49\linewidth]{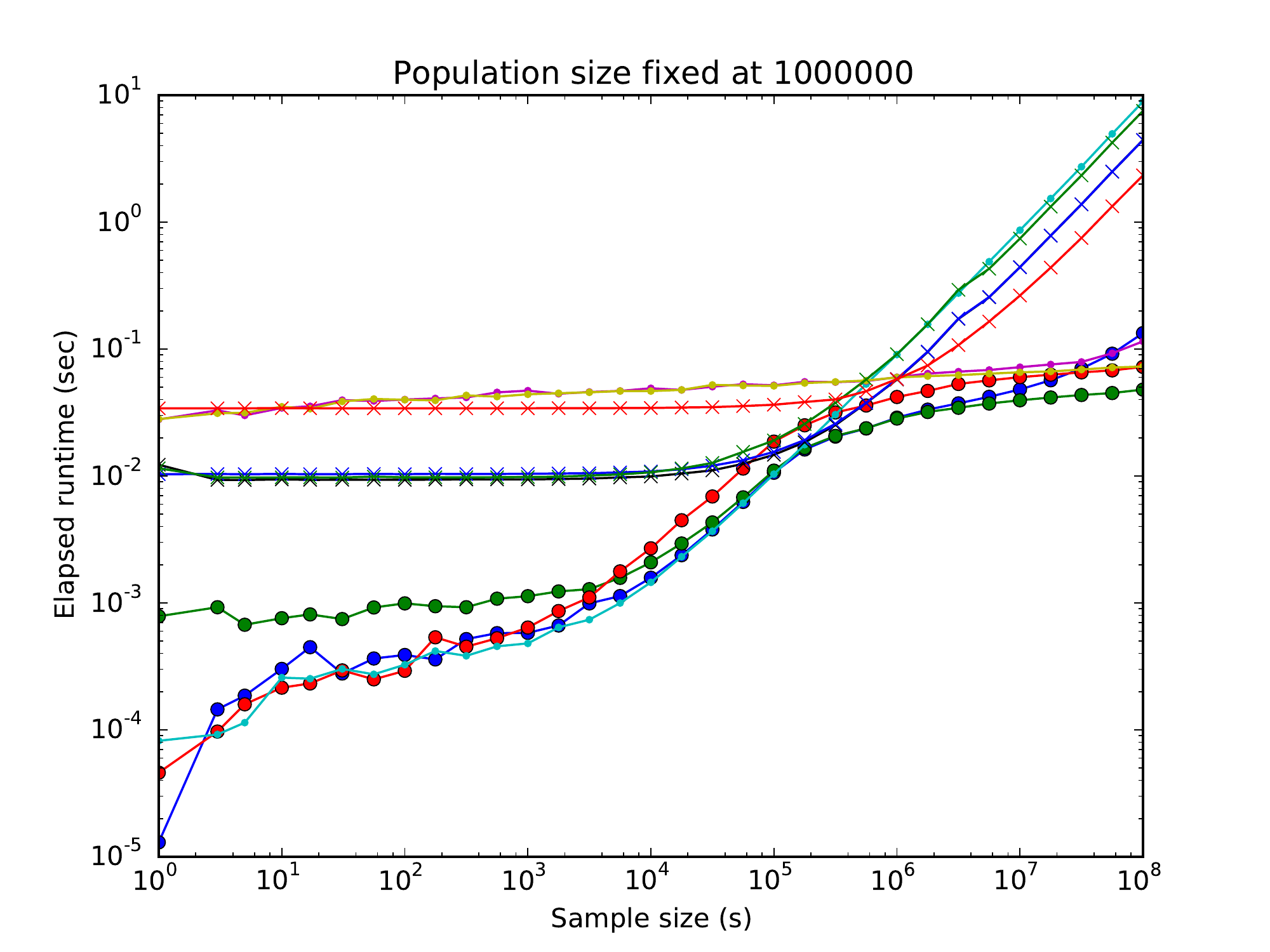} & \includegraphics[width=0.49\linewidth]{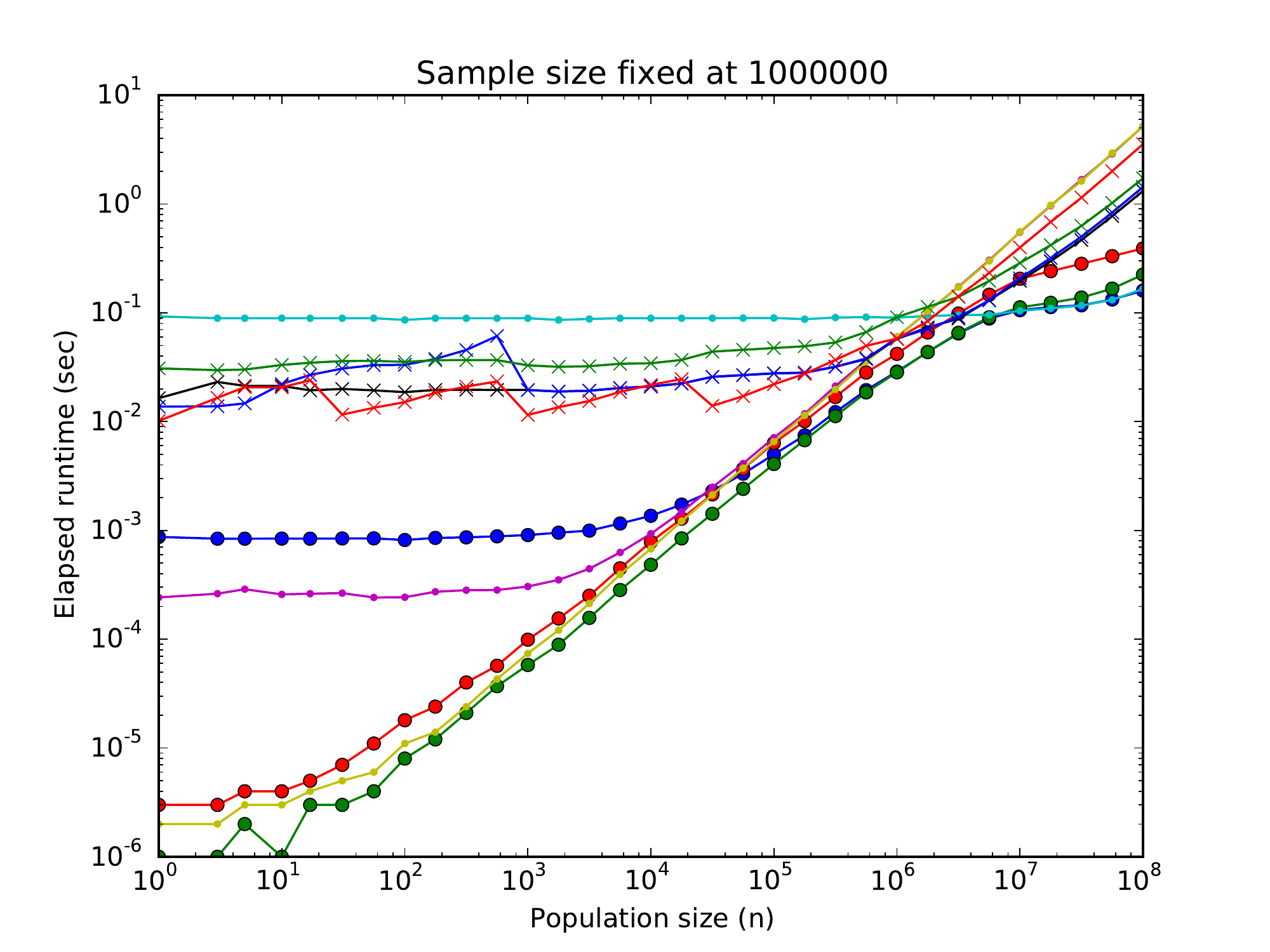} \\
\hline
\multicolumn{2}{|c|}{"Uniform" population} \\
\hline
\includegraphics[width=0.45\linewidth]{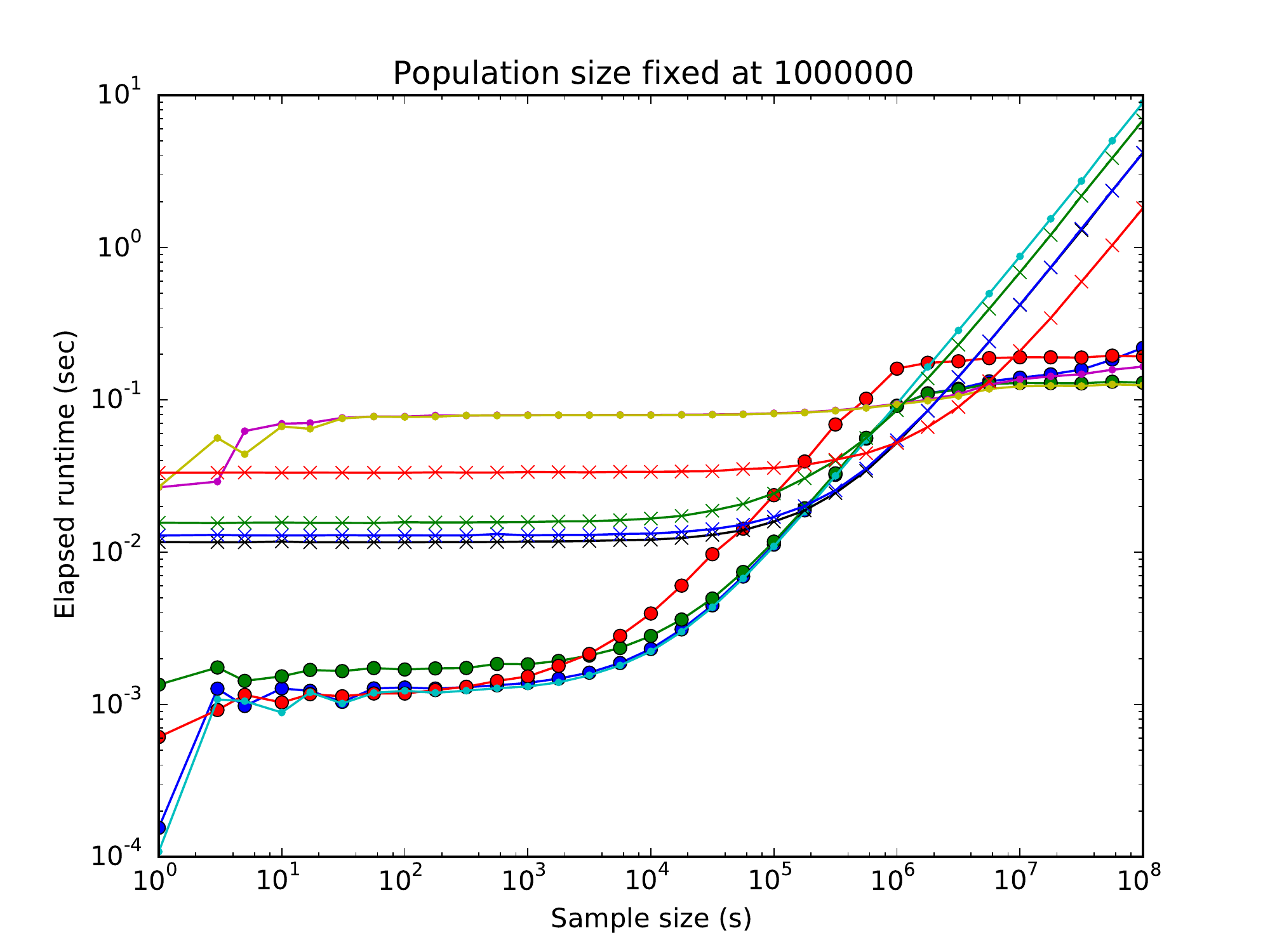} & \includegraphics[width=0.45\linewidth]{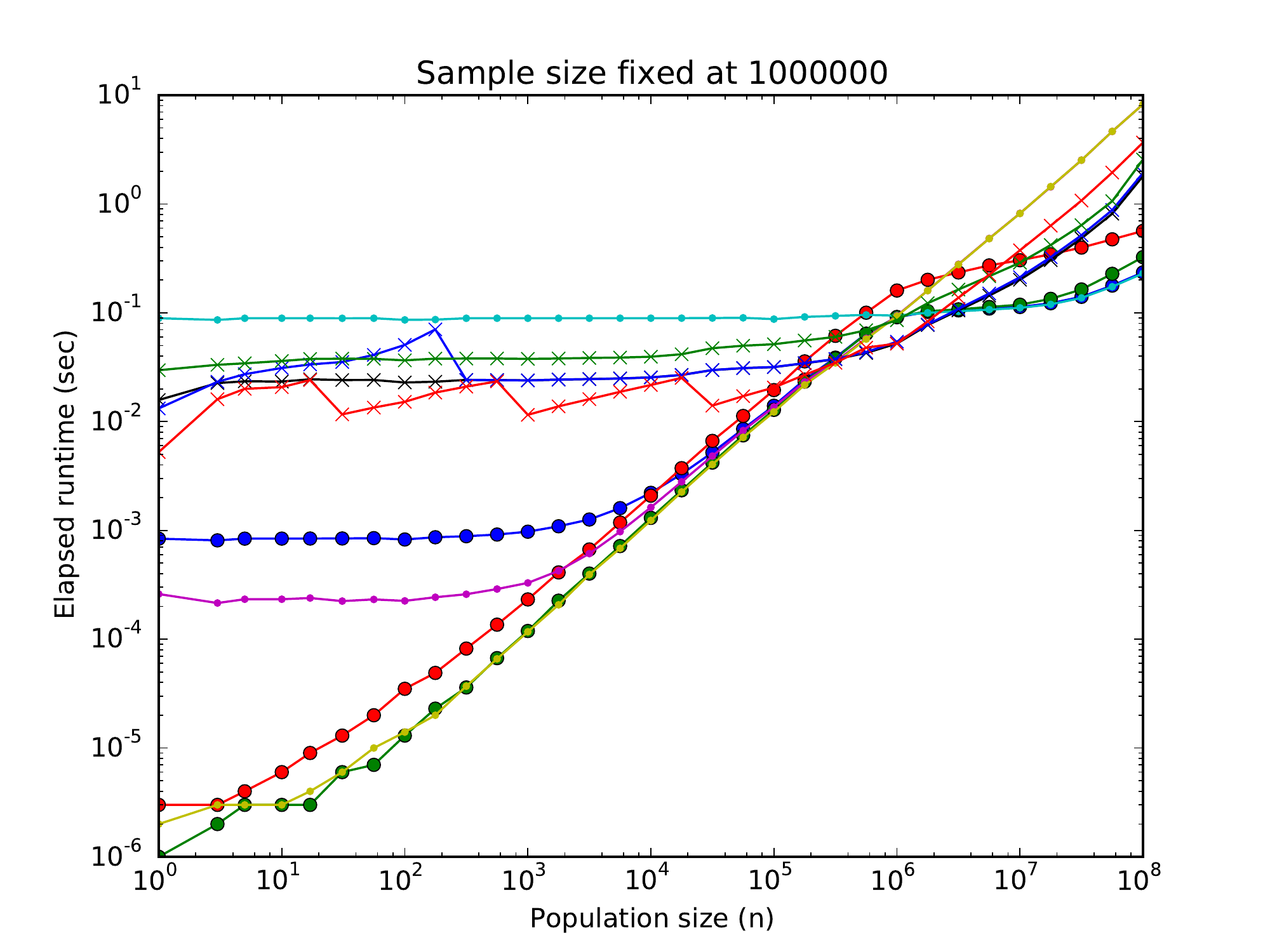} \\
\hline
\multicolumn{2}{|c|}{"Geometric" population} \\
\hline
\includegraphics[width=0.45\linewidth]{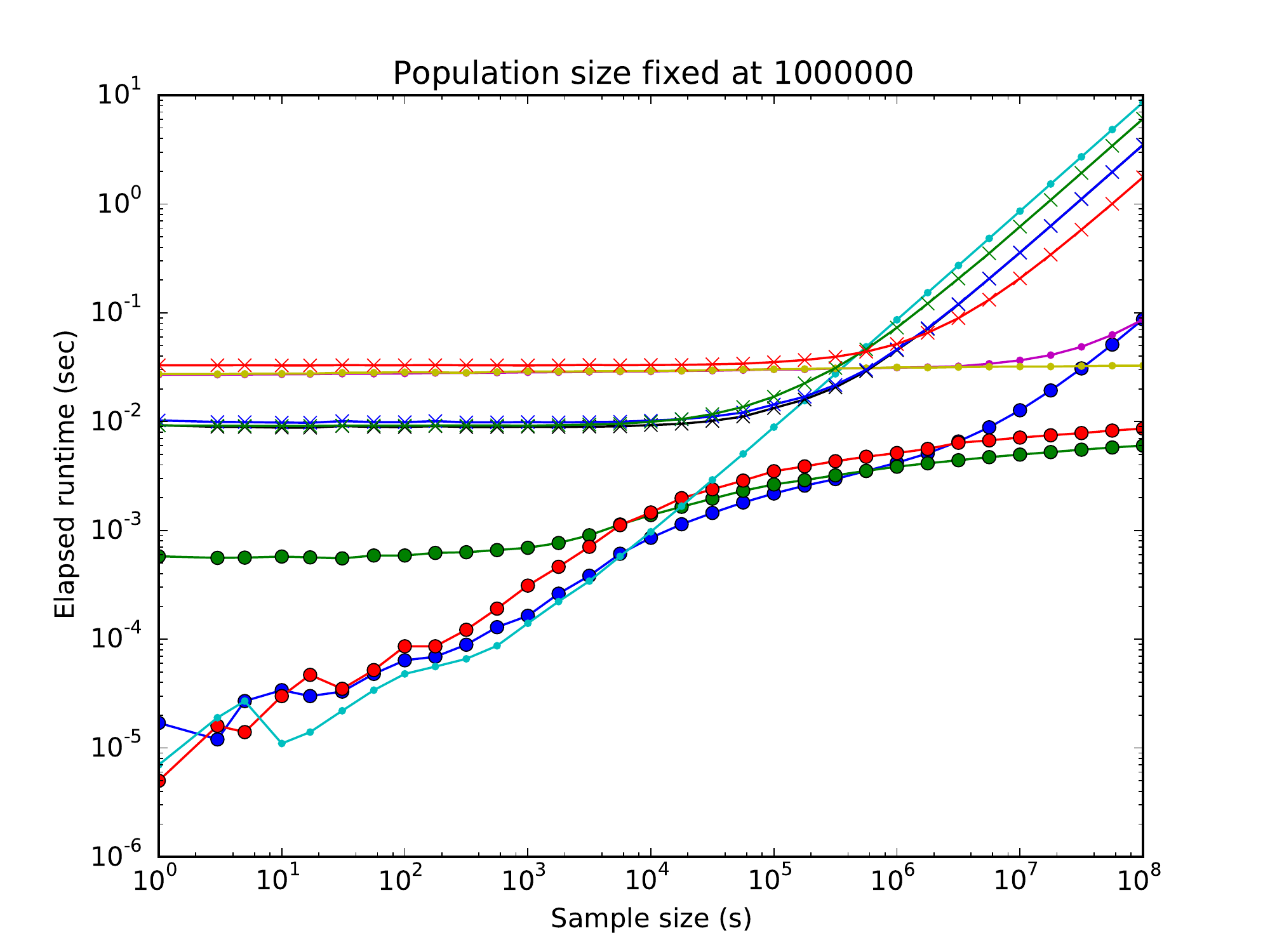} & \includegraphics[width=0.45\linewidth]{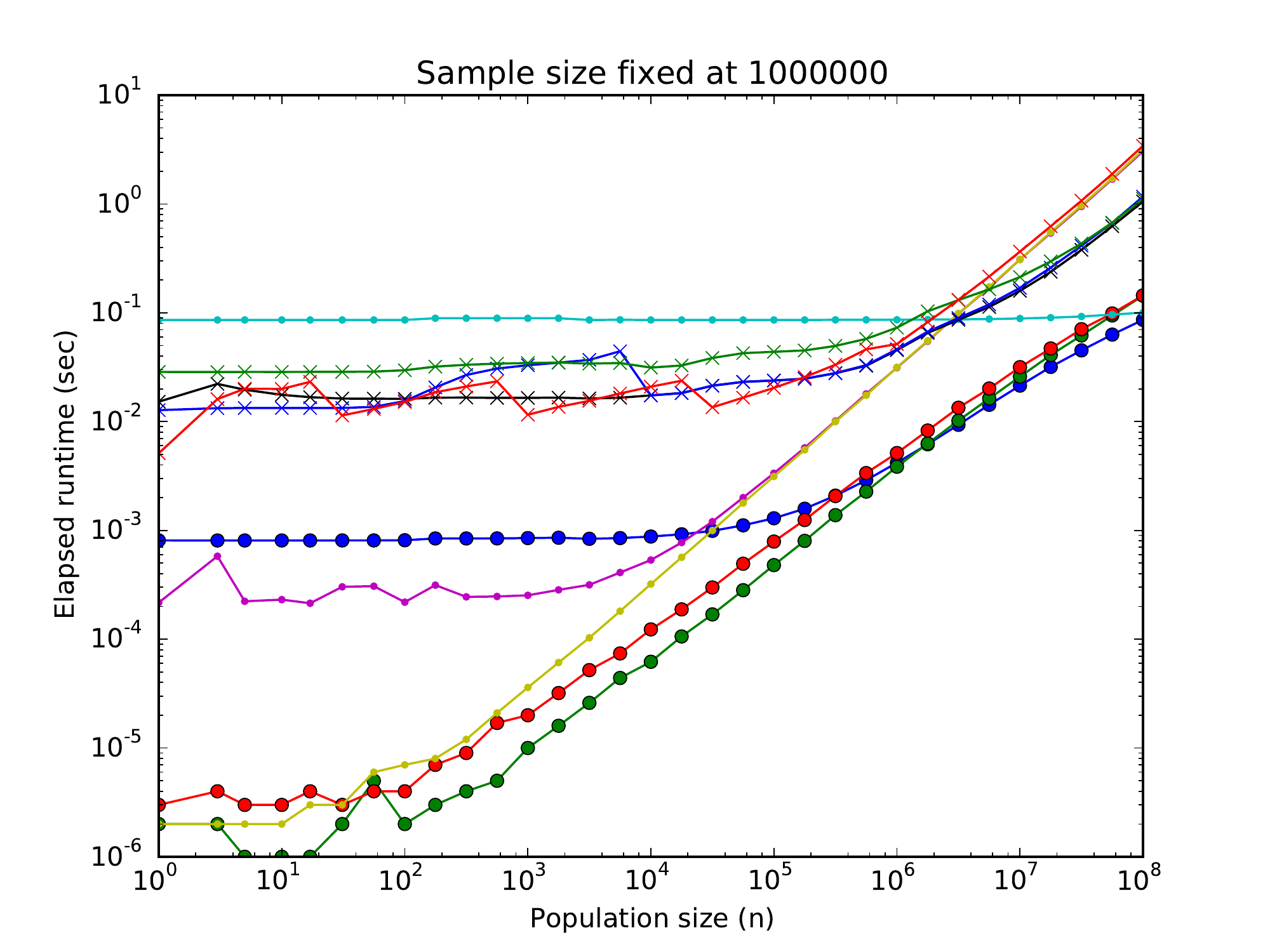} \\
\hline

\end{tabular}
\caption{Comparison of runtimes of various random choice algorithms. The various versions of the novel algorithm proposed in this work are marked with large dots, some example algorithms are marked with small dots, competing
algorithms used by various scientific libraries are marked with crosses. }
\label{fig-res}
\end{figure}

The algorithms presented in this paper have been implemented in C++11 programming language for purposes of testing and speed comparison. These are compared with the implementation of 
standard Walker's alias method~\cite{walker} as implemented by the R programming language (later referred to under name "Walker-R"), the full implementation of R's sampling function
(which examines the data, heuristically chooses between Walker's algorithm or a naive algorithm, and the runs it), referred to as "full-R", an alternative, standalone
implementation of Walker's algorithm in C by ransampl library, and with numpy's implementation which follows Algorithm~\ref{alg:one}. 
The novel algorithm proposed in this work (Algorithm~\ref{alg:five}) has been
tested in two versions, one which produces an array of size $s$ (the sample with repetitions), 
referred to as the "Alg5-array" algorithm, and one which produces the multiset (with integer counts instead of the repetitions). The multiset was implemented either 
as a hashtable (using standard C++ \emph{unordered\_map} data structure) or trivially as an array of size $n$. The implementation using the former is referred to as "Alg5-HT" in
plots, the one using the latter as "Alg5-Pop".
Similarly, Algorithm~\ref{alg:four} has been implemented and tested both outputting an array with repetitions and multiset (based on an array, hashtable implementation was produced but 
skipped in effort to avoid complicating the plots further). A summary of theoretical properties of each of the algorithms is presented in Table~\ref{tbl-the}

The R functions are written in C code, and the code for these functions is reused, slightly modified to remove dependencies on R's internals. Numpy, although written in Python, is 
compiled to native code using Cython, and as such, should run at near-native speeds, like the rest of the tested algorithms, without suffering overhead due to Python being an
interpreted, high-level language. 

The algorithms have been tested on three different random populations, one drawn from uniform distribution $\mathcal{U}([0, 1])^n$, representing population with mostly equal probabilities,
one drawn from a geometric sequence starting with $1.0$ and ending at $10^{-100}$, representing a population with skewed probabilities. The third type of population is generated
by applying a Gaussian PDF function to $n$ points evenly spaced between $0.0$ and 10 times the stdev of the Gaussian function. This is meant to simulate the usual application
of sampling function in modelling in population genetics (which in fact was the inspiration for this research): in population genetics models, selection and reproduction
of modelled organisms is often done precisely by randomly sampling with replacement of $n$ organisms (that reproduce and pass their offspring to a next generation) from a population 
of $n$. The probability of a given organism being chosen to reproduce is proportional to its \emph{fitness function} - which is often Gaussian. 

Each population type (uniform, geometric and Gaussian) is rescaled 
so that it sums to $1.0$, and randomly shuffled. The results of the tests are presented on Figure~\ref{fig-res}.

It is evident from the results of the tests that not only is the proposed algorithm asymptotically optimal, but, unlike Algorithm 4 it is also efficient in practice, outperforming the competing methods in most scenarios,
by as much as several orders of magnitude in some cases. In the single pessimistic case, where the distribution of probabilities in the population is close to uniform and $s \approx n$, although it runs slightly slower, it still remains 
competetive, moreover, the difference in runtime grows smaller as $s = n \to \infty$, and it overtakes the Walker's method at $s = n \approx 10^8$ (data not shown). 

The algorithm is able to adapt to the input data and use any skew from uniform distribution to its advantage, to increase its runtime, as evidenced by the tests on Gaussian and especially geometric populations.
Unlike the popular algorithms it works in constant additional memory, and is capable of online operation.

An implementation of this algorithm in a few programming languages may be downloaded from \url{http://bioputer.mimuw.edu.pl/~mist/stats}

\section{Application: mass sampling from any discrete distribution}

As the proposed algorithm is online, it may accept an infinite sequence of states as its population, and can still be expected to produce a sample in finite time, without exhausting the 
whole sequence. As such, one application is immediately obvoius: mass sampling of iid variates from any discrete distribution. 
All one needs ito do is to exhaustively walk through the configuration space
of the distribution, preferably (though not necessarily) in order of decreasing probability mass function (PMF), and feed the resulting sequence into the proposed algorithm. The result
is a sample from the input distribution of any desired size. 

The advantage of the proposed solution is that the input distribution does not need to have an easily invertible CDF, only a computable PMF. The runtime is usually sublinear, 
wrt. to the sample size, but that depends on the exact properties of the distribution being sampled, distributions with light tails being faster to sample from than heavy-tailed ones. 
As an example: generating a sample of size $10^9$ from Poisson distribution with $\lambda = 10000.0$ using R programming language's \emph{rpois} function takes about 90 seconds, while 
using the scheme proposed above elapses 0.7 seconds. Such an algorithm itself consumes constant memory plus any memory needed for datastructures needed to walk through the configuration
space (trivially constant in case of distributions with integer support, at worst a linear "visited" hashtable plus a linear priority queue when the configuration space is complicated, and
needs to be traversed in a Dijkstra-like fashion~\cite{dijkstra}). 
The algorithm works online, in the meaning that the generated part of the sample is immediately available for consumption, before computations proceed
to generate the rest of it.

This could be used to provide an alternative implementation of sampling functions in many programming languages, most of which
accept an argument denoting sample size, but then proceed to generate even a large sample in naive, iterative fashion. One point worth noting, however, is that the algorithm, as presented,
returns the sample sorted in the order in which the configuration space was traversed. If this is undesirable, a Fisher-Yates shuffle~\cite{shuffle} may be performed on the resulting stream,
at the cost of loss of online property.

\section{Acknowledgments}
I would like to thank prof. Anna Gambin, B{\l}a{\.z}ej Miasojedow PhD, and Mateusz {\L}{\k{a}}cki MSc for
their helpful comments.
This research was funded by grant no. 2012/06/M/ST6/00438 by Polish National Science Centre, and grant POLONIUM ,,Matematyczne i obliczeniowe modelowanie ewolucji ruchomych element{\'o}w genetycznych''.

\bibliographystyle{acm}

\medskip

\end{document}